\documentclass[onecolumn,showpacs,11.5pt]{revtex4-1}
\usepackage{graphicx,amsmath}
\usepackage{dcolumn}
\usepackage{bm}
\begin{document}
\newcommand{\hs}{\hspace*{0.5cm}}
\newcommand{\vs}{\vspace*{0.5cm}}
\newcommand{\be}{\begin{equation}}
\newcommand{\ee}{\end{equation}}
\newcommand{\bea}{\begin{eqnarray}}
\newcommand{\eea}{\end{eqnarray}}
\newcommand{\ben}{\begin{enumerate}}
\newcommand{\een}{\end{enumerate}}
\newcommand{\bde}{\begin{widetext}}
\newcommand{\ede}{\end{widetext}}
\newcommand{\nn}{\nonumber}
\newcommand{\crn}{\nonumber \\}
\newcommand{\Tr}{\mathrm{Tr}}
\newcommand{\non}{\nonumber}
\newcommand{\noi}{\noindent}
\newcommand{\al}{\alpha}
\newcommand{\la}{\lambda}
\newcommand{\bet}{\beta}
\newcommand{\ga}{\gamma}
\newcommand{\va}{\varphi}
\newcommand{\om}{\omega}
\newcommand{\pa}{\partial}
\newcommand{\+}{\dagger}
\newcommand{\fr}{\frac}
\newcommand{\bc}{\begin{center}}
\newcommand{\ec}{\end{center}}
\newcommand{\Ga}{\Gamma}
\newcommand{\de}{\delta}
\newcommand{\De}{\Delta}
\newcommand{\ep}{\epsilon}
\newcommand{\varep}{\varepsilon}
\newcommand{\ka}{\kappa}
\newcommand{\La}{\Lambda}
\newcommand{\si}{\sigma}
\newcommand{\Si}{\Sigma}
\newcommand{\ta}{\tau}
\newcommand{\up}{\upsilon}
\newcommand{\Up}{\Upsilon}
\newcommand{\ze}{\zeta}
\newcommand{\ps}{\psi}
\newcommand{\Ps}{\Psi}
\newcommand{\ph}{\phi}
\newcommand{\vph}{\varphi}
\newcommand{\Ph}{\Phi}
\newcommand{\Om}{\Omega}

\title{ \bf
Electroweak sphalerons in the reduced minimal 3-3-1 model}
\author{Vo Quoc  Phong\footnote{vqphong@hcmus.edu.vn}}
\affiliation{  Department of Theoretical Physics, Ho Chi Minh City
University of Science, Ho Chi Minh City, Vietnam}
\author{Hoang Ngoc Long\footnote{hnlong@iop.vast.ac.vn}}
\affiliation{Institute of Physics, Vietnamese  Academy of Science and Technology,
10 Dao Tan, Ba Dinh, Hanoi, Vietnam}
\author{Vo Thanh Van\footnote{vtvan@hcmus.edu.vn}}
\affiliation{Department of Theoretical Physics, Ho Chi Minh City
University of Science, Ho Chi Minh City, Vietnam}
\author{Nguyen Chi Thanh\footnote{ncthanh.phs@gmail.com}}
\affiliation{Department of Theoretical Physics, Ho Chi Minh City
University of Science, Ho Chi Minh City, Vietnam}
\date{\today}

\begin{abstract}
We calculate the electroweak sphaleron rates in the reduced
minimal 3-3-1 (RM331) model. In the context of the early Universe,
this model undergoes a sequence of two first-order phase
transitions, $SU(3) \rightarrow SU(2)$ at the TeV scale and $SU(2)
\rightarrow U(1)$ at the $10^2$ GeV scale, as the Universe cools
down from the hot big bang. By a thin-wall approximation, we show
that for each phase transition in this sequence, the sphaleron
rate is larger than the cosmological expansion rate at temperatures higher than the critical temperature, and
after the phase transition, the sphaleron process is decoupled.
This may provide baryon-number violation (B violation) necessary
for baryogenesis in the relationship with nonequilibrium physics in
the early Universe.
\end{abstract}

\maketitle


\section{Introduction}\label{secInt} 
Electroweak baryogenesis (EWBG) is a possibility to explain the baryon asymmetry of the Universe (BAU) by electroweak physics. From an initially baryon-antibaryon-symmetric Universe, the nonzero BAU can be generated if the three Sakharov conditions are satisfied: B violation, C and CP violations, and deviation from thermal equilibrium \cite{sakharov}. Cohen, Kaplan and Nelson \cite{ckn} have proposed an EWBG mechanism in which the presence of B violation and CP violation may be related to each other in a nonequilibrium way which would produce a BAU. This mechanism requires that the expanding Universe experience a first-order phase transition period in which bubbles of broken electroweak symmetry nucleate, grow, collide and merge in the midst of symmetric phase regions. B violation will happen quickly in the symmetric phase regions, but it will shut off essentially in the broken phase bubbles; this gives a relationship between out-of-equilibrium and B-violating processes necessary for baryogenesis. On the other hand, CP violation will come from the interactions between fermions, which become massive through electroweak spontaneous symmetry breaking (SSB), and the bubble walls; this provides a relationship between CP violation and nonequilibrium physics during cosmological expansion.

When the Universe cools through the phase-transition critical temperature $T_c$, the electroweak phase transition (EWPT) associated with SSB takes place. In the symmetric phase, the Higgs potential has only one minimum at place where the vacuum expectation value (VEV) of the Higgs field is zero. As the temperature reaches the critical temperature, the Higgs fields will tend to get a nonzero VEV in a manifold of equivalent vacua, and the new minima of the Higgs potential appear \cite{vilenkin}. If the EWPT is of the first order, there exists a potential barrier which separates the new minima from the old minimum of the Higgs potential, and the transition will occur through bubble nucleation. At this transitional time, if the temperature is small as compared to the height of the potential barrier, the transition may occur by quantum tunneling in which the magnitude of the Higgs field changes from zero to nonzero VEV; such a transition is called an \textit{instanton}. Otherwise, if the temperature is sufficiently high so that thermal fluctuations can bring the magnitude of the Higgs field from zero VEV over the barrier to nonzero VEV classically without tunneling, the transition is called a \textit{sphaleron}.

In 1976, 't Hooft discovered the B-violation instantons \cite{thooft}, but the tunneling amplitude in the Standard Model (SM) is too small for baryogenesis. In 1984, Klinkhamer and Manton \cite{manton} found the sphaleron as a static, saddle-point solution of the classical field equations in the SM; and in 1985, Kuzmin, Rubakov, and Shaposhnikov \cite{shaposhnikov} showed that at the temperature $T\geq 100$ GeV, the B-violation sphalerons can take place with significant probabilities and the sphaleron rate is larger than the cosmological expansion rate.

The B-violation sphalerons have been investigated in the SM and various extended models \cite{mssm, sphatwohiggs}. In the SM, the sphaleron rate is very small, about $10^{-60}$ \cite{manton, sphasm, Farrar, Arnold, spha-nonpur}; this rate is much smaller than the rate of BAU and smaller than the cosmological expansion rate. In the extended models, the sphaleron processes have been considered with various suppositions, but the results show that the B-violation rates are small in the symmetric phase and smaller than the Universe's expansion rate \cite{decoupling}.

Among the extended models, those based on the $\mathrm{SU}(3)_C\otimes \mathrm{SU}(3)_L \otimes \mathrm{U}(1)_X$ gauge group (called 3-3-1 for short) \cite{ppf, flt, ecn331, rm331} have some intriguing features, such as the ability to account for the generation problem \cite{ppf, flt} or the quantization of the electric charge \cite{chargequan}. We hope that the 3-3-1 models can also answer the BAU problem. In the present work, we investigate the electroweak sphalerons in the RM331 model \cite{rm331} because of its simplicity. This model consists of the minimal leptonic content (i.e., only the SM leptons) and bileptons: the singly and the doubly charged gauge bosons $V^{\pm}$ and $U^{\pm\pm}$, the heavy neutral boson $Z_2$ and the exotic quarks. This model also has two Higgs triplets. Therefore, the physical scalar spectrum of the RM331 model is composed of a doubly charged scalar $h^{++}$ and two neutral scalars $h_1$ and $h_2$ \cite{rm331}. These new particles and exotic quarks can be triggers for the first-order phase transition.

This paper is organized as follows: In Sec. \ref{sec2}, we give a review of the EWPT in the RM331 model. In Sec. \ref{sec3}, we present the sphaleron energy for calculation of the sphaleron rate. In Sec. \ref{sec4}, we investigate the sphaleron rates of the phase transitions $SU(3) \rightarrow SU(2)$ and $SU(2) \rightarrow U(1)$. Finally,  we summarize and describe outlooks in Sec. \ref{sec5}.

\section{A review of the EWPT in the RM331 model}\label{sec2}

In our previous work \cite{phonglongvan}, we have used an effective potential at finite temperature to study the structure of the EWPT in the RM331 model. In order to derive that effective potential, we start from the full Higgs Lagrangian
\begin{equation} \label{HiggsLagrangian}
\mathcal{L}=\left( \mathcal{D}_{\mu }\chi \right) ^{\dagger }\left( \mathcal{%
D}^{\mu }\chi \right) +\left( \mathcal{D}_{\mu }\rho \right)
^{\dagger }\left( \mathcal{D}^{\mu }\rho \right)-V(\chi,\rho),
\end{equation}
where
\begin{equation} \label{HiggsPotential}
V(\chi,\rho)=\mu^2_1\rho^\dagger\rho+ \mu^2_2\chi^\dagger\chi+\la_1(\rho^\dagger\rho)^2+\la_2(\chi^\dagger\chi)^2
+\la_3(\rho^\dagger\rho)(\chi^\dagger\chi)+\la_4(\rho^\dagger\chi)(\chi^\dagger\rho).
\end{equation}

Expanding  $\rho$ and $\chi$ around  $v_{\rho}$ and $v_{\chi}$, which are considered as variables \footnote{At $0^K$, $v_{\rho}\equiv v_{\rho_0}=246$ GeV and $v_{\chi}\equiv v_{\chi_0}=4 \div 5$ TeV; in this work, we choose $v_{\chi_0}=4$ TeV}, we obtain
\begin{equation}
\mathcal{L}=\frac{1}{2}\partial^{\mu}v_\chi\partial_{\mu}v_\chi+
\frac{1}{2}\partial^{\mu}v_\rho\partial_{\mu}v_\rho-V_0(v_{\chi}, v_{\rho})
+\sum_{boson}m^2_{boson}(v_{\chi}, v_{\rho})W^{\mu}W_{\mu},\label{l1}
\end{equation}
where $W$ runs over all gauge fields and Higgs bosons. We can split the masses of particles into two parts as follows:
\begin{equation}
m^2_{boson}(v_{\chi}, v_{\rho})=
 m^2_{boson}(v_{\chi})+m^2_{boson}(v_{\rho}).
\end{equation}

The RM331 model has the following gauge bosons: two massive bosons like the SM bosons $Z_1$ and  $W^{\pm}$, the new heavy neutral boson $Z_2$, the singly and doubly charged gauge bosons $U^{\pm\pm}$ and $V^{\pm}$, two doubly charged Higgses $h^{++}$ and  $h^{--}$, one heavy neutral Higgs $h_2$, and one SM-like Higgs $h_1$. The masses of the gauge bosons and the Higgses in the RM331 model are presented in Table \ref{tab1}.

\begin{table}
\caption{Mass formulations of bosons in the RM331 model.}
\bc
\begin{tabular}{|l|l|c|c|}
\hline
Bosons &$m^2(v_\chi,v_\rho)$ & $ m^2(v_\chi)$  & $m^2(v_\rho)$ \\
\hline
$m_{W^{\pm }}^{2}$ &$\frac{g^{2}v_{\rho }^{2}}{4}$ & 0 &$80.39^2$ $(\mathrm{GeV})^2$ \\
\hline
$m_{V^{\pm }}^{2}  $&$\frac{g^{2}v_{\chi }^{2}}{4}$ &$1307.15^2$ $(\mathrm{GeV})^2$ &0\\
\hline
$m_{U^{\pm \pm }}^{2} $&$\frac{g^{2}\left( v_{\rho }^{2}+
v_{\chi }^{2}\right) }{4}$&$1307.15^2$ $(\mathrm{GeV})^2$ &$80.39^2$ $(\mathrm{GeV})^2$ \\
\hline
$m^2_{Z_1}\sim m^2_{Z} $ &$\frac{1}{4}\frac{g^2}{\cos^2\theta_W}
 v_{\rho}^2$ &0 &$91.682^2$ $(\mathrm{GeV})^2$ \\
\hline
$m^2_{Z_2}\sim m^2_{Z'}$ &$\frac{1}{3} \: g^2 \left[\frac{\cos^2\theta_W}{1\!-\!4
\sin^2\theta_W} \: v_{\chi}^2 + \:  \frac{1\!- \!4
\sin^2\theta_W}{4\!\cos^2\theta_W}\:v_{\rho}^2\right]$ &$4.8^2$
$(\mathrm{TeV})^2$ & $14.53^2$ $(\mathrm{GeV})^2$ \\
\hline
$m^2_{h_1} $&$ \left(\la_1 -\frac{\la^2_3}{4\la_2}\right)v^2_\rho$ &0&
$125^2$ $(\mathrm{GeV})^2$\\
\hline
$m^2_{h^{++}} $&$ \frac{\lambda_4}{2}(v^2_\chi + v^2_\rho)$
&$\frac{\lambda_4}{2}v^2_\chi$ &$\frac{\lambda_4}{2} v^2_\rho$\\
\hline
$m^2_{h_2}$&$\la_2v^2_\chi +\frac{\la^2_3}{4\la_2}v^2_\rho$ &
$\la_2v^2_\chi $ &$\frac{\la^2_3}{4\la_2}v^2_\rho$ \\
\hline
\end{tabular}
\ec
\label{tab1}
\end{table}

The structure of the EWPT in the RM331 model is divided into two parts, $SU(3) \rightarrow SU(2)$ and $SU(2)\rightarrow U(1)$ \cite{phonglongvan}. Due to the fact that the two scales of symmetry breaking are much different, $v_{\chi_0} \gg  v_{\rho_0}$ ($v_{\chi_0}\sim 4-5$ TeV \cite{rm331, dkck}, $v_{\rho_0}=246$ GeV) and that the Universe is accelerating, the SSBs can take place sequentially, in which the symmetry breaking $SU(3)\rightarrow SU(2)$ takes place before the symmetry breaking $SU(2)\rightarrow U(1)$.

Through the boson mass formulations in Table \ref{tab1}, the boson $V^{\pm}$ is only involved in the first phase transition $SU(3) \rightarrow SU(2)$. The gauge bosons $Z_1$, $W^{\pm}$ and $h_1$ are only involved in the second phase transition, $SU(2) \rightarrow U(1)$. However, $U^{\pm\pm}$, $Z_2$ and $h^{--}$ are involved in both phase transitions. The reason why $U^{\pm\pm}$, for example, can get mass in both phase transitions is as follows: When the energy of the Universe lowers to the scale $v_{\chi_0}$, the symmetry breaking $SU(3)\rightarrow SU(2)$ generates mass for the first part of $U^{\pm\pm}$; i.e., $U^{\pm\pm}$ eats one of the Goldstone bosons $\chi^{\pm\pm}$ of the triplet $\chi$. As the Universe cools to the scale $v_{\rho_0}$, the symmetry breaking $SU(2)\rightarrow U(1)$ is turned on, which generates mass for the last part of  $U^{\pm\pm}$; i.e., $U^{\pm\pm}$ eats another of the Goldstone bosons $\rho^{\pm\pm}$ of the triplet $\rho$.

The symmetry breaking $SU(3)\rightarrow SU(2)$ through $\chi_0$ generates masses for the heavy gauge bosons — such as  $U^{\pm\pm}$, $V^{\pm}$, $Z_2$ — and the exotic quarks. The symmetry breaking scale, $v_{\chi_0}$, is chosen to be $4$ TeV \cite{rm331, dkck}. This phase transition involves exotic quarks and heavy bosons,  without the involvement of the SM particles. Therefore, the effective potential \cite{phonglongvan} can be written as
\begin{equation} \label{EP-SU3}
V_{SU(3) \rightarrow SU(2)}^{eff}=D'(T^2-{T'}^2_0){v_{\chi}}^2-E'
Tv_{\chi}^3 +\frac{\lambda'_T}{4}v_{\chi}^4,
\end{equation}
where
\bea \label{thhh}
D'&=& \frac{1}{24 {v_{\chi_0}}^2} \left\{6 m_U^2(v_{\chi})+ 3m_{Z_2}^2(v_{\chi})+6 m_{V}^2(v_{\chi})
+ 18m_Q^2(v_{\chi})+ m_{h_2}^2(v_{\chi})+2 m_{h^{\pm\pm}}^2(v_{\chi}) \right\},\crn
{T'}_0^2 &=&  \frac{1}{D}\left\{\frac{1}{4} m_{h_2}^2(v_{\chi}) -
 \frac{1}{32\pi^2v_{\chi_0}^2} \left(6 m_U^4(v_{\chi})+ 3 m_{Z_2}^4(v_{\chi})+6 m_{V}^4(v_{\chi}) -
 36 m_Q^4(v_{\chi}) \right.\right.\crn
&&\qquad \left.\left.+m_{h_2}^4(v_{\chi})+ 2 m_{h^{\pm\pm}}^4(v_{\chi})\right)\right\},\crn
E' &=& \frac{1}{12 \pi v_{\chi_0}^3} (6 m_U^3(v_{\chi}) + 3 m_{Z_2}^3(v_{\chi}) +
6 m_{V}^3(v_{\chi}) +m_{h_2}^3(v_{\chi})+ 2 m_{h^{\pm\pm}}^3(v_{\chi})),\crn
\lambda'_T &=&
 \frac{m_{h_2}^2(v_{\chi})}{2 v_{\chi_0}^2} \left\{ 1 - \frac{1}{8\pi^2 v_{\chi_0}^2 m_{h_2}^2(v_{\chi})}
 \left[6 m_V^4(v_{\chi}) \ln \frac{m_V^2(v_{\chi})}{b T^2} +
 3 m_{Z_2}^4(v_{\chi}) \ln \frac{m_{Z_2}^2(v_{\chi})}{b T^2}
 \right.\right.\crn
&&\qquad \left.\left. +6 m_U^4(v_{\chi}) \ln \frac{m_U^2(v_{\chi})}{bT^2}-36 m_Q^4(v_{\chi}) \ln
\frac{m_Q^2(v_{\chi})}{b_F T^2} +m_{h_2}^4(v_{\chi}) \ln \frac{m_{h_2}^2(v_{\chi})}{b
T^2}\right.\right.\crn
&&\qquad \left.\left.+ 2  m_{h^{\pm\pm}}^4(v_{\chi}) \ln \frac{m_{h^{\pm\pm}}^2(v_{\chi})}{b
T^2} \right] \right\}.\eea

In Eq. \eqref{thhh}, $T'_0$ is the temperature at which the phase transition $SU(3)\rightarrow SU(2)$ ends. As the temperature drops below $T'_0$, the minimum of effective potential \eqref{EP-SU3} at $v_{\chi}=0$ disappears, and the gauge symmetry $\mathrm{SU}(3)_L \otimes \mathrm{U}(1)_X$ is totally broken. Once $T'_0$ is specified, we can calculate the temperature $T'_1$, above which the symmetry $\mathrm{SU}(3)_L \otimes \mathrm{U}(1)_X$ is restored and below which the phase transition starts (and the bubble nucleation occurs, in the case of the first-order phase transition) \cite{mkn}:
\begin{equation} \label{T'1}
T'_1=\frac{T'_0}{\sqrt{1-\frac{9E'^2}{8D'\lambda'_{T'_c}}}}.
\end{equation}

At the symmetry-breaking scale $v_{\chi_0}$, the phase-transition strength is decided by the masses of the heavy neutral Higgs $h_2$, the doubly charged Higgses $h^{++}$ and the exotic quarks. In order to have a first-order phase transition, this strength must be larger than unity; i.e., $\frac{v_{\chi_c}}{T'_c}=\frac{2E'}{\lambda'_{T'_c}}\geq 1$, where $T'_c$ is the critical temperature, at which the values of $V_{eff}(v_{\chi})$ at the minima become equal:
\be \label{T'c}
T'_c=\frac{T'_0}{\sqrt{1-E'^2/D'\lambda'_{T'_c}}}. 
\ee And the heavy particle masses must be in the range of a few TeV \cite{phonglongvan}.

When the Universe has been expanding and cooling down to the scale $v_{\rho_0}$, the phase transition $SU(2)\rightarrow U(1)$ is turned on through $\rho_0$, which generates the masses of the SM particles and the last part of the mass of  $U^{\pm\pm}$. With the symmetry-breaking scale equal to $ v_{\rho_0}= $ $v_0=246$ GeV, the high-temperature expansion of the effective potential has the form \cite{phonglongvan}
\begin{equation} \label{V_eff_rho}
V_{eff}^{RM331}(v_\rho)=D(T^2-T^2_0).v^2_{\rho}-ET|v_{\rho}|^3+\frac{\lambda_T}{4}v^4_{\rho},
\end{equation}
where
\bea
D &=& \frac{1}{24 {v_0}^2} \left[6 m_W^2(v_{\rho})+6 m_U^2(v_{\rho})
+3m_{Z_1}^2(v_{\rho}) +3 m_{Z_2}^2(v_{\rho}) \right.\crn
&&\quad \left.+ 6 m_t^2 (v_{\rho})+m_{h_1}^2(v_{\rho})+m_{h_2}^2(v_{\rho})
 + 2 m_{h^{\pm\pm}}^2(v_{\rho}) \right],\crn
T_0^2 &=&  \frac{1}{D}\left\{\frac{1}{4} (m_{h_1}^2(v_{\rho})+m_{h_2}^2(v_{\rho})) -
 \frac{1}{32\pi^2v_0^2} \left(6 m_W^4(v_{\rho})+6 m_U^4(v_{\rho})
 + 3 m_{Z_1}^4(v_{\rho}) \right.\right.\crn
&&\qquad \left.\left.
 +3 m_{Z_2}^4(v_{\rho}) - 12 m_t^4(v_{\rho}) + m_{h_1}^4(v_{\rho})
 +m_{h_2}^4(v_{\rho}) + 2 m_{h^{\pm\pm}}^4(v_{\rho})\right)
\right\},\crn
E &=& \frac{1}{12 \pi v_0^3} \left(6 m_W^3(v_{\rho})+6 m_U^3(v_{\rho})
+ 3 m_{Z_1}^3(v_{\rho}) +3 m_{Z_2}^3(v_{\rho}) \right.\crn
&&\quad \left. +m_{h_1}^3(v_{\rho})+m_{h_2}^3(v_{\rho})
+ 2 m_{h^{\pm\pm}}^3(v_{\rho})\right),\label{trilinear}\\
\lambda_T &=&
 \frac{m_{h_1}^2(v_{\rho})+ m_{h_2}^2(v_{\rho})}{2 v_0^2}
 \left\{ 1
- \frac{1}{8\pi^2 v_0^2 (m_{h_1}^2(v_{\rho})+m_{h_2}^2(v_{\rho}))}
 \left[6 m_W^4(v_{\rho}) \ln \frac{m_W^2(v_{\rho})}{b T^2} \right.\right.\crn
&&\qquad \left.\left. + 3 m_{Z_1}^4(v_{\rho}) \ln\frac{m_{Z_1}^2(v_{\rho})}{b T^2}
+3 m_{Z_2}^4(v_{\rho}) \ln \frac{m_{Z_2}^2(v_{\rho})}{b T^2}+6 m_U^4(v_{\rho})
 \ln \frac{m_U^2(v_{\rho})}{bT^2}\right.\right.\crn
&&\qquad \left.\left.-12 m_t^4(v_{\rho}) \ln\frac{m_t^2(v_{\rho})}{b_F T^2}
+ m_{h_1}^4(v_{\rho}) \ln \frac{m_{h_1}^2(v_{\rho})}{b T^2}
+m_{h_2}^4(v_{\rho}) \ln \frac{m_{h_2}^2(v_{\rho})}{b T^2}\right.\right.\crn
&&\qquad  \qquad  \qquad  \qquad  \left.\left. + 2m_{h^{\pm\pm}}^4(v_{\rho})
\ln \frac{m_{h^{\pm\pm}}^2(v_{\rho})}{b T^2} \right] \right\}\nn \eea

In the limit $E\rightarrow 0$, we have a second-order phase transition. In order to have a first-order phase transition, the phase-transition strength has to be larger than unity; i.e., $\frac{v_{\rho_c}}{T_c}=\frac{2 E}{\lambda_{T_c}}\geq 1$, where the critical temperature $T_c$ is given by
\be \label{Tc}
T_c=\frac{T_0}{\sqrt{1-E^2/D\lambda_{T_c}}}.
\ee
We also have the bubble nucleation temperature $T_1$:
\begin{equation} \label{T1}
T_1=\frac{T_0}{\sqrt{1-\frac{9E^2}{8D\lambda_{T_c}}}}.
\end{equation}
Note that $T_0$ in Eqs. \eqref{trilinear}, \eqref{Tc}, and \eqref{T1} is the transition-ending temperature of the EWPT $SU(2)\rightarrow U(1)$.  

From Eqs. \eqref{V_eff_rho} and \eqref{trilinear}, the effective potential $V_{eff}^{RM331}(v_\rho)$ depends on the masses of the SM particles and a part of the masses of the new particles, $h_2$ and $h^{\pm\pm}$. If we forget all contributions of these new particles, the EWPT $SU(2)\rightarrow U(1)$ of the RM331 model becomes that of the SM; in this case, however, with $m_{h_1}=125$ GeV, the phase-transition strength cannot be larger than unity, and then no first-order phase transition can exist. Therefore, these new particles act as triggers for the first-order phase transition.

The mass regions of $h_1$ and $h^{++}$ for the first-order phase transition are \cite{phonglongvan}
\begin{equation} 200 \,
\mathrm{GeV}   < m_{h^{++}}(v_\rho) < 1200\,  \mathrm{GeV} \end{equation}
and
\begin{equation} 0  <  m_{h_2}(v_\rho) < 624 \,  \mathrm{GeV}.\end{equation}

In order to calculate the sphaleron energies, we also choose the mass of $h_1$ and $h^{++}$ in these regions.

\section{Sphaleron energy}\label{sec3}

The RM331 model is a type of non-Abelian gauge theory which incorporates the Higgs mechanism. In such a theory, the vacuum has a nontrivial structure which has the degenerated minima separated by energy barriers in the field configuration space \cite{mkn}. For the different minima, we have the different baryon and lepton numbers. As a consequence, each transition between these different minima is accompanied by a change in the baryon number. And the B violation can be seen throughout the sphaleron processes \cite{sphalerons}.

In order to study the sphaleron processes in the RM331 model, we consider the Lagrangian of the gauge-Higgs system:
\begin{equation}\label{gauge-Higgs}
\mathcal{L}_{\rm gauge-Higgs} =-\frac{1}{4}F^{a}_{\mu\nu}F^{a\mu\nu}
    + \left( \mathcal{D}_{\mu }\chi \right) ^{\dagger }\left( \mathcal{%
D}^{\mu }\chi \right) +\left( \mathcal{D}_{\mu }\rho \right)
^{\dagger }\left( \mathcal{D}^{\mu }\rho \right)-V(\chi,\rho).
\end{equation}

From Eq.~(\ref{gauge-Higgs}), the energy functional in the temporal gauge takes the form
\begin{equation} \label{Energy-01}
\mathcal{E} = \int d^3\boldsymbol{x}
    \bigg[\left( \mathcal{D}_{\mu }\chi \right) ^{\dagger }\left( \mathcal{%
D}^{\mu }\chi \right) +\left( \mathcal{D}_{\mu }\rho \right)
^{\dagger }\left( \mathcal{D}^{\mu }\rho
\right)+V(\chi,\rho)\bigg];
\end{equation}
here we assume that the least energy has the pure-gauge configurations, hence $F^a_{ij}=0$.

By the temperature expansion of the effective potential from the previous section, the energy functional is reduced to
\begin{equation}\label{Energy-02}
\mathcal{E} = 4\pi\int^\infty_0d^3x\bigg[\frac{1}{2}\bigg(\nabla^2 v_\chi\bigg)^2
+\frac{1}{2}\bigg(\nabla^2 v_\rho\bigg)^2
    +V_{\rm eff}(v_\chi, v_\rho;T)\bigg].
\end{equation}

Using the static field approximation as follows,
\begin{equation}\label{StaticApprox}
\frac{\partial v_\chi}{\partial t}=\frac{\partial v_\rho}{\partial t}=0,
\end{equation}
we obtain
\begin{equation}\label{Energy-03}
\mathcal{E}=\int d^3x \left[ \frac{1}{2}(\partial_i v_{\chi})^2+\frac{1}{2}(\partial_i v_{\rho})^2+
V_{eff}(v_\chi, v_\rho;T)\right].
\end{equation}

From the Lagrangian (\ref{l1}), we have two equations of motion for the VEVs:
\begin{subequations}\label{cd}
\begin{equation}\label{cda}
\ddot{v_\chi}+\nabla^2 v_\chi-\frac{\partial V_{eff}(v_\chi,T)}{\partial v_\chi} = 0
 \end{equation}
and
 \begin{equation} \label{cdb}
\ddot{v_\rho}+\nabla^2 v_\rho-\frac{\partial V_{eff}(v_\rho,T)}{\partial v_\rho}=0.
\end{equation}
\end{subequations}

By the static field approximation \eqref{StaticApprox}, we rewrite Eq. \eqref{cd} in spherical coordinates:
\begin{subequations}\label{4.161}
\begin{equation}\label{4.161a}
\frac{d^2v_\chi}{dr^2}+\frac{2}{r}\frac{dv_\chi}{dr}-\frac{\partial V_{eff}(v_\chi,T)}{\partial v_\chi}=0
 \end{equation}
and
 \begin{equation} \label{4.161b}
\frac{d^2v_\rho}{dr^2}+\frac{2}{r}\frac{dv_\rho}{dr}-\frac{\partial V_{eff}(v_\rho,T)}{\partial v_\rho}=0.
\end{equation}
\end{subequations}

For the RM331 model, the EWPT $SU(3)\rightarrow SU(2)$ takes place as the temperature drops to a few TeV, while the EWPT $SU(2)\rightarrow U(1)$ occurs as the temperature is about 100 GeV. From the energy functional \eqref{Energy-03}, we have the sphaleron energies in each phase transition as follows:
\begin{subequations}\label{4.163}
\begin{equation}\label{4.163a}
 \mathcal{E}_{sph.su(3)}=4\pi\int\bigg[\frac{1}{2}\left(\frac{dv_\chi}{dr}\right)^2+V_{eff}(v_\chi,T)\bigg]r^2dr
 \end{equation}
and
 \begin{equation} \label{4.163b}
\mathcal{E}_{sph.su(2)}=4\pi\int\bigg[\frac{1}{2}\left(\frac{dv_\rho}{dr}\right)^2+V_{eff}(v_\rho,T)\bigg]r^2dr,
\end{equation}
\end{subequations}
where $ \mathcal{E}_{sph.su(3)}$ and $\mathcal{E}_{sph.su(2)}$ are the sphaleron energies in the $SU(3)\rightarrow SU(2)$ and $SU(2)\rightarrow U(1)$ phase transitions, respectively.

In order to calculate these energies, we must solve the equations of motion \eqref{4.161} for the VEVs of the Higgs fields.

\section{Sphaleron rate}\label{sec4}

The sphaleron rate per unit time per unit volume, $\Gamma/V$, is characterized by a Boltzmann factor, $\exp\left(-\mathcal{E}/T\right)$, as follows \cite{ctspha, gauge0}:
\begin{equation} \label{SphaleRateDef}
\Gamma/V = \alpha^4 T^4 \exp\left(-\mathcal{E}/T\right),
\end{equation}
where $V$ is the volume of the EWPT's region,  $T$ is the temperature, $\mathcal{E}$ is the sphaleron energy, and $\alpha=1/30$. We will compare the sphaleron rate with the Hubble constant, which describes the cosmological expansion rate at the temperature $T$ \cite{cross, spha-huble}:
\begin{equation} \label{Hubble}
H^2=\frac{\pi^2 g T^4}{90 M_{pl}^2},
\end{equation}
where  $g=106.75$, $M_{pl}=2.43 \times 10^{18}$ GeV.

In order to have B violation, the sphaleron rate must be larger than the Hubble rate at the temperatures above the critical temperature (otherwise, B violation will become negligible during the Universe's expansion); however, the sphaleron process must be decoupled after the EWPT to ensure the generated BAU is not washed out \cite{decoupling}.

\subsection{Upper bounds of the sphaleron rates}

To estimate the upper bounds of the sphaleron rates, we suppose that the VEVs of the Higgs fields do not change from point to point in the Universe. Due to this supposition, we have $\frac{dv_\chi}{dr}=\frac{dv_\rho}{dr}=0$. Hence, from Eq. \eqref{4.161} we obtain
\begin{equation}\label{dkk}
\frac{\partial V_{eff}(v_\chi)}{\partial v_\chi}=0, \quad \frac{\partial V_{eff}(v_\rho)}{\partial v_\rho}=0.
\end{equation}

Equation (\ref{dkk}) shows that $v_\chi$ and $v_\rho$ are the extremes of the effective potentials. Therefore, the sphaleron energies \eqref{4.163} can be rewritten as
\begin{subequations}\label{4.165}
\begin{equation}\label{4.165a}
 \mathcal{E}_{sph.su(3)}=4\pi\int V_{eff}(v_\chi,T) r^2dr=\frac{4\pi r^3}{3}V_{eff}(v_\chi,T) \bigg|_{v_{\chi_m}}
 \end{equation}
and
 \begin{equation} \label{4.165b}
 \mathcal{E}_{sph.su(2)}=4\pi\int V_{eff}(v_\rho,T) r^2dr=\frac{4\pi r^3}{3}V_{eff}(v_\rho,T) \bigg|_{v_{\rho_m}},
 \end{equation}
\end{subequations}
where $v_{\chi_m}, v_{\rho_m}$ are the VEVs at the maximum of the effective potentials. From Eq. \eqref{4.165}, the sphaleron  energies are equal to the maximum heights of the potential barriers.

The Universe's volume  at a temperature $T$ is given by $V=\frac{4\pi r^3}{3}=\frac{1}{T^3}$. Because the whole Universe is an identically thermal bath, the sphaleron energies are approximately
\begin{equation}\label{4.20}
 \mathcal{E}_{sph.su(3)}\sim \frac{{E'}^4 T}{4{\lambda'}_T^3}, \quad \mathcal{E}_{sph.su(2)}\sim \frac{E^4 T}{4\lambda^3_{T}}.
\end{equation}

From the definition \eqref{SphaleRateDef}, the sphaleron rates take the forms
\begin{subequations}\label{4.21}
\begin{equation}\label{4.21a}
\Gamma_{su(3)}=\alpha_w^4 T\exp\left({-\frac{{E'}^4T}{4{\lambda'}_{T}^3T}}\right) 
 \end{equation}
and
 \begin{equation} \label{4.21b}
\Gamma_{su(2)}=\alpha_w^4 T\exp\left({-\frac{E^4T}{4\lambda_{T}^3T}}\right).
 \end{equation}
\end{subequations}

\begin{figure}[h]
\centering
\includegraphics[height=9cm,width=16cm]{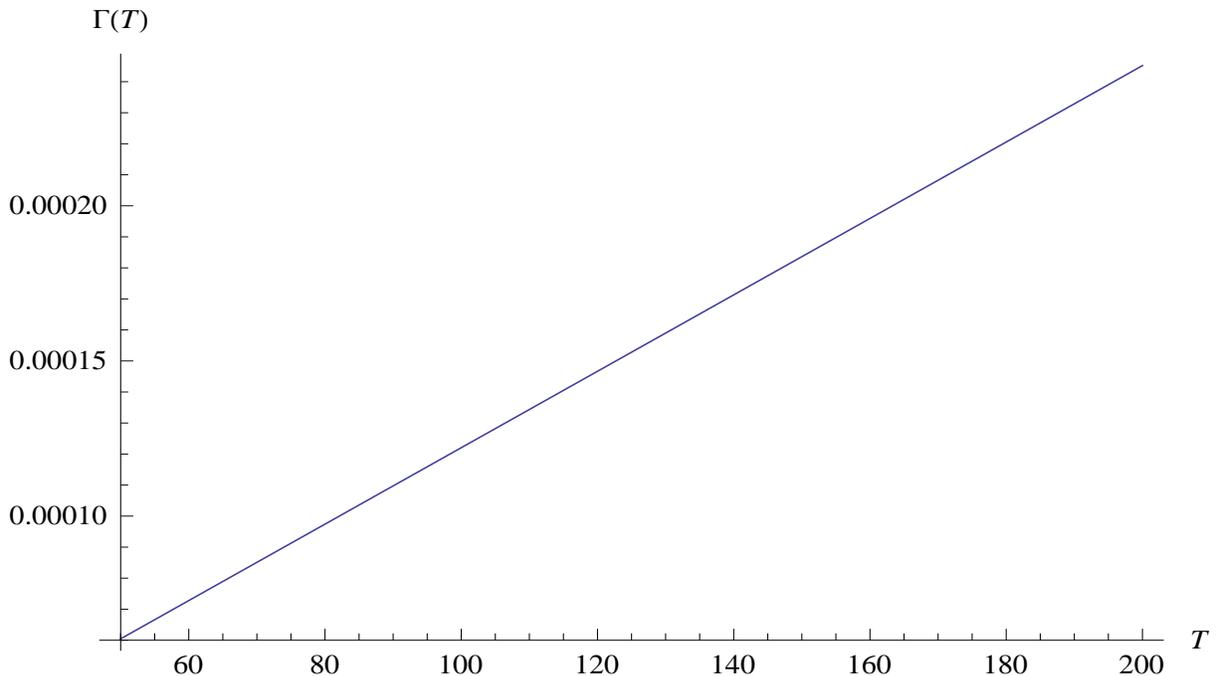}
\caption{The sphaleron rate in the phase transition $SU(2)\rightarrow U(1)$.
We choose $m_{h_2}(v_\rho)=100$ GeV, $m_{h^{\pm\pm}}(v_\rho)=350$ GeV.}\label{fig:02}
\end{figure}

For the heavy particles, $E, \lambda, E'$ and $\lambda'$ are constant. Hence, the sphaleron rates in this approximation are the linear functions of temperature, as illustrated in Fig. \ref{fig:02} for the case of the phase transition $SU(2)\rightarrow U(1)$.

From Eq. \eqref{4.21}, we estimate the upper bounds of the sphaleron rates as follows:
\begin{equation}\label{UpBound}
\Gamma_{su(3)}\sim 10^{-3}\gg H; \quad \Gamma_{su(2)} \sim 10^{-4} \gg  H \sim 10^{-13}.
\end{equation}

In this approximation, however, the sphaleron decoupling condition cannot be satisfied. For instance, with $m_{h_2}(v_\rho)=100$ GeV and $m_{h^{\pm\pm}}({v_\rho})=350$ GeV, as the temperature drops below the phase-transition temperature $T_c=138.562$ GeV and the Universe switches to the symmetry-breaking phase, the sphaleron rate is still much larger than the Hubble constant, and this makes the B violation washed out. By this consequence, the sphaleron process cannot occur identically in large regions of space; it can only take place in the microscopic regions.

\subsection{Sphaleron rates in a thin-wall approximation}

\begin{figure}[h]
\centering
\includegraphics[height=15cm,width=10cm]{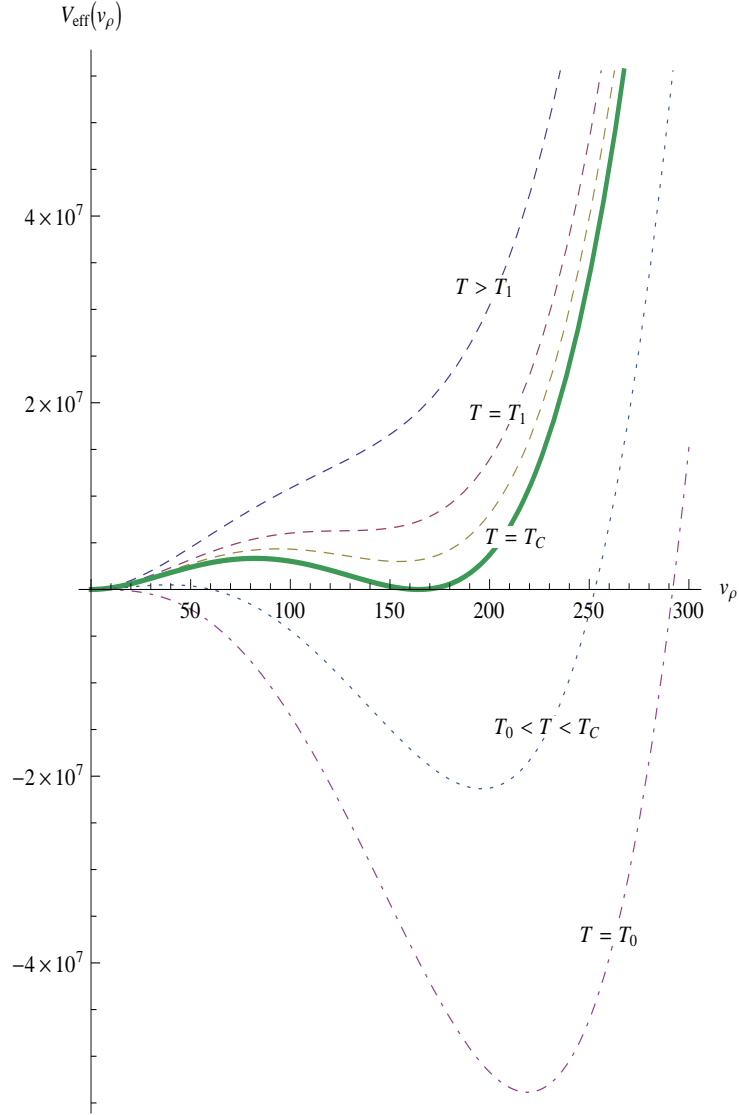}
\caption{The effective potential $V_{eff}^{RM331}(v_\rho)$ in the EWPT $SU (2)\rightarrow U(1)$. We choose $m_{h_2}(v_\rho)=100$ GeV, $m_{h^{\pm\pm}}(v_\rho)=350$ GeV. We obtain the bubble nucleation temperature $T_1\approx 141.574$ GeV, the critical temperature $T_c=138.562$ GeV,  and the transition-ending temperature $T_0=118.42$ GeV.}\label{fig:03}
\end{figure}

At every point in the early Universe, the effective potential varies as a function of magnitude of the Higgs field at various temperatures, as illustrated in Fig. \ref{fig:03} for the case of $V_{eff}^{RM331}(v_\rho)$. If the temperature at a spatial location is higher than the bubble nucleation temperature $T_1$, then $V_{eff}^{RM331}(v_\rho)$ at this location has only one minimum at $v_\rho=0$, and this location belongs to a symmetric phase region. As the temperature drops below $T_1$, the second minimum of $V_{eff}^{RM331}(v_\rho)$ gradually forms, and the potential barrier which separates two minima gradually appears. At this spatial location, $v_\rho$ can be changed by thermal fluctuations so that $V_{eff}^{RM331}(v_\rho)$ gets the second minimum. The phase transition occurs microscopically, resulting in a tiny bubble of broken phase in which the Higgs field $\rho$ acquires a nonzero expectation value.  As the temperature reaches the critical temperature $T_c$,  the second minimum becomes equal to the first minimum of $V_{eff}^{RM331}(v_\rho)$. But when the temperature goes below $T_c$, the second minimum becomes the lower one corresponding to a true vacuum, while the first minimum becomes the false vacuum. Such tiny true-vacuum bubbles at various locations in the Universe can occur randomly and expand in the midst of false vacuum. If the sphaleron rate is larger than the Universe's expansion rate, the bubbles can collide and merge until the true vacuum fills all space. However, if the sphaleron decoupling condition is satisfied after the transition, the sphaleron rate must be smaller than the cosmological expansion rate when the temperature goes from $T_c$ to $T_0$, at which the first minimum of $V_{eff}^{RM331}(v_\rho)$ completely disappears.

\begin{subequations}\label{TW}
Suppose that in a bubble of the phase transition $SU(3)\rightarrow SU(2)$ ,  we have
\begin{equation}\label{TWa}
\frac{\partial V_{eff}(v_\chi)}{\partial v_\chi}\approx \frac{\Delta V_{eff}(v_\chi)}{\Delta v_{\chi}}=const \equiv M';
 \end{equation}
here $\Delta v_\chi=v_{\chi_c}$, $\Delta V_{eff}(v_\chi)= V_{eff}(v_{\chi_c})-V_{eff}(0)$, and $v_{\chi_c}$ is a  second minimum of the effective potential for this transition. Similarly, in a bubble of the phase transition $SU(2)\rightarrow U(1)$, we have
 \begin{equation} \label{TWb}
\frac{\partial V_{eff}(v_{\rho})}{\partial v_\rho}\approx \frac{\Delta V_{eff}(v_\rho)}{\Delta v_{\rho}}=const \equiv M;
 \end{equation}
\end{subequations}
here $\Delta v_\rho=v_{\rho_c}$, $\Delta V_{eff}(v_\rho)= V_{eff}(v_{\rho_c})-V_{eff}(0)$, and $v_{\rho_c}$ is a  second minimum of the effective potential for the phase transition.

Now, we solve the equations of motion \eqref{4.161} for the VEVs $v_\chi$ and $v_\rho$ by the approximation \eqref{TW}. Rewritting Eq. \eqref{4.161} in this approximation, we have
\begin{subequations}\label{prdd}
\begin{equation}\label{prdda}
\frac{d^2v_\chi}{dr^2}+\frac{2}{r}\frac{dv_\chi}{dr} = M'
 \end{equation}
and
 \begin{equation} \label{prddb}
\frac{d^2v_\rho}{dr^2}+\frac{2}{r}\frac{dv_\rho}{dr}= M.
 \end{equation}
\end{subequations}

In the cases that $r \to \infty$ (the spatial locations are in the symmetric phase) or $r \to 0$ (the spatial locations are in the broken phase), the VEVs must satisfy the boundary conditions:
\begin{equation} \label{InLimits}
\lim_{r\rightarrow \infty}v_{\chi}(r)=\lim_{r\rightarrow \infty}v_\rho(r)=0; \quad \frac{dv_\chi (r)}{dr} \bigg|_{r=0}= \frac{dv_\rho (r)}{dr}\bigg|_{r=0}=0.
\end{equation}

In the bubble walls, the solutions of Eq. \eqref{prdd} take the forms
\begin{subequations}\label{VEV}
\begin{equation}\label{VEVa}
v_\chi=\frac{M'}{6}r^2-A'/r+B'
 \end{equation}
and
 \begin{equation} \label{VEVb}
v_\rho=\frac{M}{6}r^2-A/r+B,
 \end{equation}
\end{subequations}
where $A', B', A, B$ are the parameters to be specified.

The continuity of the scalar fields in a bubble results in the two following systems of equations. The first is the system for the EWPT $SU(3) \rightarrow SU(2)$:
\begin{subequations}\label{rs}
\begin{equation}\label{rsa}
\begin{cases}
\frac{M'}{6}R^2_{b.su(3)}-A'/R_{b.su(3)}+B'=v_{\chi_c},\\
\frac{M'}{6}(R_{b.su(3)}+\Delta l')^2-A'/(R_{b.su(3)}+\Delta l')+B'=0,
\end{cases}
\end{equation}
where $R_{b.su(3)}$ and $\Delta l'$ are, respectively, the radius and the wall thickness of a bubble which is nucleated in the phase transition $SU(3) \rightarrow SU(2)$. The second system is that for the EWPT $SU(2) \rightarrow SU(1)$:
\begin{equation}\label{rsb}
\begin{cases}
\frac{M}{6}R^2_{b.su(2)}-A/R_{b.su(2)}+B=v_{\rho_c},\\
\frac{M}{6}(R_{b.su(2)}+\Delta l)^2-A/(R_{b.su(2)}+\Delta l)+B=0,
\end{cases}
\end{equation}
where $R_{b.su(2)}$ and $\Delta l$ are, respectively, the radius and the wall thickness of a bubble nucleated in the phase transition $SU(2)\rightarrow U(1)$.
\end{subequations}

Solving the systems of Eq. \eqref{rs}, we obtain the solutions $v_\chi$ and $v_\rho$, which are of the forms
\begin{subequations}\label{tuong}
\begin{equation}\label{tuong1}
v_\chi(r)=\begin{cases}
v_{\chi_c};\quad \text{when } r \leq R_{b.su(3)}, \\
\frac{M'}{6}r^2-A'/r+B';\quad  \text{when } R_{b.su(3)}<r \leq R_{b.su(3)}+\Delta l'\\
0;\quad  \text{when } R_{b.su(3)}+\Delta l'<r
\end{cases}
\end{equation}
and
\begin{equation}\label{tuong2}
v_\rho(r)=\begin{cases}
v_{\rho_c};\quad \text{when } r \leq R_{b.su(2)}, \\
\frac{M}{6}r^2-A/r+B;\quad \text{when } R_{b.su(2)}<r \leq R_{b.su(2)}+\Delta l\\
0;\quad \text{when } R_{b.su(2)}+\Delta l<r.
\end{cases}
\end{equation}
\end{subequations}

To go on, we must overcome two obstacles. The first one is that we have only four equations in the systems \eqref{rs}, but we have to specify eight unknown parameters [$A'$, $B'$, $\Delta l'$, $R_{b.su(3)}$, $A$, $B$, $\Delta l$, and $R_{b.su(2)}$]. To overcome this, we suppose the sphaleron rate to be equal to the Hubble rate at the critical temperature. This supposition relies on the requirement for avoiding the washout of the generated BAU after a phase transition, by which the sphaleron rate must be larger than the Hubble rate at temperatures above the critical temperature, but the sphaleron rate must be smaller than the Hubble rate at temperatures below the critical temperature.

The second obstacle is that the masses of many heavy particles in the RM331 model are unknown so far. However, we can estimate their mass regions which satisfy the conditions for the first-order phase transition, and we choose any values in these mass regions for calculation of the sphaleron energies. Although the strengths of the first-order phase transitions in this model are sufficiently strong ($>1$), they are not so strong ($<5$) as shown in Ref. \cite{phonglongvan}, and hence the coefficients in the effective potential are not meaningfully different for the different values in these mass regions. Here, we choose $m_q=m_{h_2}(v_\chi)=1500$ GeV, $m_{h2}(v_\rho)=100$ GeV, and  $m_{h^{\pm\pm}}(v_\rho)=350$ GeV.

\begin{figure}[h]
\centering
\includegraphics[height=12cm,width=14cm]{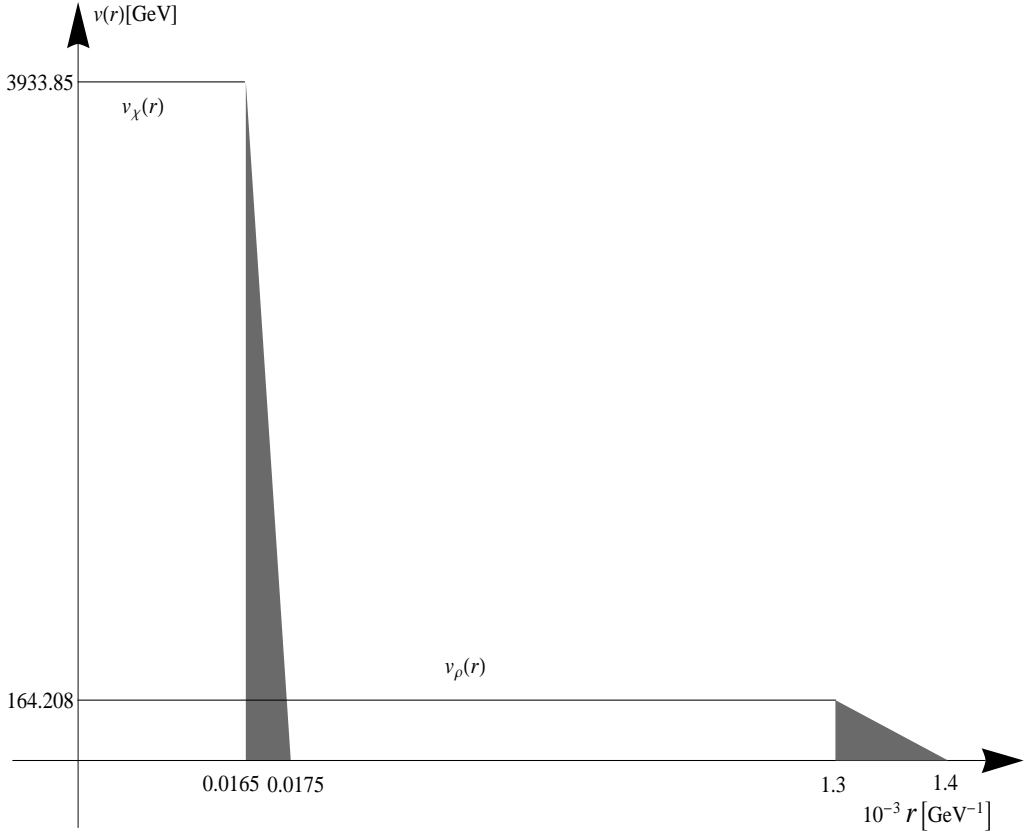}
\caption{The solutions $v_\chi(r)$ and $v_\rho(r)$ in the phase transitions $SU(3)\rightarrow SU(2)$ and $SU(2)\rightarrow U(1)$, respectively. We choose $m_{h_2}(v_\chi)=1500$ GeV, $m_{q}(v_\chi)=1500$ GeV, $m_{h_2}(v_\rho)=100$ GeV, and $m_{h^{\pm\pm}}(v_\rho)=350$ GeV. The regions in grey portray the thin walls of vacuum bubbles nucleated in each phase transition.}\label{fig:04}
\end{figure}

In Fig. \ref{fig:04}, we show our respective solutions $v_\chi(r)$ and $v_\rho(r)$. These solutions are not as smooth as those in Refs. \cite{decoupling, gauge0, momenta}. The reason is that the bubble walls we consider in this work are very thin, $\Delta l,\Delta l' \ll 1/T$ (while in Ref. \cite{momenta}, for instance, the authors consider the case in which $\Delta l \gg 1/T$). Inside the thin walls of the bubbles, the derivatives $\frac{d v_\chi}{dr}$ and $\frac{d v_\rho}{dr}$ are very large; this allows the Higgs fields $\chi$ and $\rho$ to change their values over the potential barriers. Therefore, the thinner the bubble walls, the larger the sphaleron rates.

The calculation results for the sphaleron rates of both phase transitions in the sequence of EWPTs are shown in Tables \ref{tab.02} and \ref{tab.03}. These results show the behavior of the sphaleron rates in the cosmological expansion as the Universe cools through the respective critical temperature of each transition in the sequence.

\begin{table}
\caption{The sphaleron rate in the EWPT $SU(3)\rightarrow SU(2)$ with $m_q(v_\chi) = m_{h2}(v_\chi) = 1500\,  {\rm GeV}$.}
\begin{tabular}{l|l|c|c|c|c|c}
\hline\hline
$T$              & $R_{b.su(3)}$  & $R_{b.su(3)}/\Delta l'$  & $\mathcal{E}_{sph.SU(3)}$ & $\Gamma_{SU(3)}$& $H$   & $\Gamma_{SU(3)}/H$         \\
$[GeV]$ &$[10^{-6}\times GeV^{-1}]$ & &$[GeV]$ &$[10^{-11}\times GeV]$ & $[10^{-12}\times GeV]$ &\\
\hline
1479.48 ($T'_1$)  &$10$      &$10$         &$6975.17$            &$1.63719 \times 10^{6}$ &$3.08195$ &$5.31 \times 10^6$\\
\hline
1450                      &$12$       &$12$         &$12481.3$           &$3.2702 \times 10^{4}$    &$2.96034$ &$1.10 \times 10^5$\\
\hline
1400                      &$13$       &$13$         &$17206.3$           &$7.94481 \times 10^2$      & $2.7597$ &$2.878 \times 10^3$\\
\hline
1390                      &$15$      & $15$         & $23251.7$           &$9.3264$   &$2.72042$ & $3.42$ \\
\hline
1388.4556 ($T'_c$) &$16.5$  & $16.5$       & $28135.1$          & $0.2714$   &$2.71438$  & 1 \\
\hline
1387                      &$17$      & $17$         & $29854.0$          & $0.07687$  & $2.70869$ & $0.28$ \\
\hline
1000                     &$19$       &$19$          & $60590.8$          &$5.98\times 10^{-19}$ &$1.40801$ &$4.25 \times 10^{-18}$\\
\hline
900                      & $22$       &$22$          & $89250.8$         &$9.50\times 10^{-36}$ & $1.14049$ & $8.33 \times 10^{-35}$\\
\hline
865.024 ($T'_0$)    &$25$        &$25$          &$119110.36$      &$1.69 \times 10^{-52}$ & $1.05357$ & $1.60 \times 10^{-51}$\\
\hline\hline
\end{tabular}\label{tab.02}
\end{table}

From the results in Table \ref{tab.02}, the gauge symmetry $\mathrm{SU}(3)_L \otimes \mathrm{U}(1)_X$ starts to be broken at the bubble nucleation temperature $T'_1\approx1479.48$ GeV. The tiny bubbles with radius $10^{-5}\,  {\rm GeV}^{-1}$ appear and store the nonzero $v_{\chi_0}$ inside. At this temperature, the sphaleron rate is as large as $1.63719 \times 10^{-5}$ GeV, which is $5.31 \times 10^6$ times larger than the cosmological expansion rate ($H=3.08195 \times 10^{-12}$ GeV). As the temperature drops from the nucleation temperature $T'_1$ to the critical temperature $T'_c$, the bubbles increase in size and the sphaleron rate decreases, but it is still much larger than the Hubble rate. This allows the bubbles to collide and coalesce. When the temperature reaches the critical temperature $T'_c=1388.4556$ GeV, the sphaleron rate is equal to the Hubble rate as we supposed. At temperatures below $T'_c$, the sphaleron rate decreases very quickly, and it becomes much smaller than the Hubble rate. As the temperature reaches the transition-ending temperature $T'_0=865.024$ GeV, only the broken phase remains, and the sphaleron $SU(3)\rightarrow SU(2)$ is totally shut off .

\begin{table}
\caption{The sphaleron rate in the EWPT $SU(2)\rightarrow U(1)$ with $m_{h2}(v_\rho)=100$ GeV, $m_{h^{\pm\pm}}(v_\rho)=350$ GeV.}
\begin{tabular}{l|l|c|c|c|c|c}
\hline\hline
$T$                      & $R_{s.su(2)}$  & $ R_{s.su(2)}/\Delta l$  & $\mathcal{E}_{sph.SU(2)}$ & $\Gamma_{SU(2)}$ & $H$ & $\Gamma_{SU(2)}/H$                                           \\
$[GeV]$ &$[10^{-4}\times GeV^{-1}]$ & &$[GeV]$ &$[10^{-12}\times GeV]$ & $[10^{-14}\times GeV]$ &\\
\hline
141.574 ($T_1$) &$6$              & $10$          & $742.838$       &$919936.07$         & $2.82211$  &$3.25\times 10^{7}$\\
\hline
141.5                   &$8$              & $10$          & $1020.87$       & $128525.28$         & $2.81916$  &$4.55 \times 10^6$\\
\hline
141                      &$10$            &$10$           &$1442.75$        &$6264.89$             & $2.79927$   & $2.23 \times 10^5$\\
\hline
140                      & $12$           &  $12$         & $2342.21$        &$9.37289$             &$2.7597$     & $339.6$\\
\hline
138.562 ($T_c$)  &$13.1$          & $13$         & $3135.75$        & $0.02703$            &$2.703$       &1    \\
\hline
137                      &$14$            & $14$         & $3922.29$         & $0.0000622$         & $2.6427$    & $2.357 \times 10^{-3} $\\
\hline
130                      &$16$          & $16$          &$6567.08$          &$1.847 \times 10^{-14}$ & $2.379$ & $7.76 \times 10^{-13}$\\
\hline
120                      & $18$         &$18$          &$10068.2$          &$5.403 \times 10^{-29}$ &$2.02754$ &$2.66 \times 10^{-27}$\\
\hline
118.42 ($T_0$)    &$20$          &$20$          &$12656.7$           &$5.595 \times 10^{-39}$ &$6.209$ &$9.01 \times 10^{-38}$\\
\hline\hline
\end{tabular}\label{tab.03}
\end{table}

When the Universe lowers its energy to the scale $v_{\rho_0}$ due to its expansion, a similar process takes place for the gauge symmetry $\mathrm{SU}(2)_L \otimes \mathrm{U}(1)_Y$. From Table \ref{tab.03}, the broken phase of the EWPT $SU(2)\rightarrow U(1)$ starts at the bubble nucleation temperature $T_1\approx141.574$ GeV in the bubbles with radius $6 \times 10^{-4}\,  {\rm GeV}^{-1}$. At this temperature, the sphaleron rate is $919936.07 \times 10^{-12}$ GeV, which is $3.25\times 10^7$ times larger than the Hubble rate ($H=2.82211 \times 10^{-14}$ GeV). As the temperature drops below $T_1$, the sphaleron rate is larger than the Hubble rate, and this lasts until the temperature reaches the critical temperature $T_c=138.562$ GeV. As the temperature goes from $T_c$ to $T_0$, the sphaleron rate is smaller than the Hubble rate, and it becomes negligible at $T_0=118.42$ GeV when the transition $SU(2)\rightarrow SU(1)$ ends.

For both EWPTs, baryon violation strongly takes place in regions of electroweak symmetries; however, it quickly shuts off in bubbles of broken phases due to the large Higgs VEVs. This may provide a relationship between nonequilibrium physics and baryon violation necessary for baryogenesis, according to the mechanism of Cohen, Kaplan, and Nelson \cite{ckn}. Therefore, the electroweak sphaleron in the RM331 model gives us a possibility for BAU from an initially baryon-symmetric Universe.

\section{Conclusion and outlooks}\label{sec5}

We have investigated the electroweak sphalerons in the reduced minimal 3-3-1 (RM331) model through calculating the sphaleron rates by the thin-wall approximation \eqref{TW}. For both transitions in the sequence of EWPTs in this model, $SU (3)\rightarrow SU(2)$ and $SU (2)\rightarrow U(1)$, the sphaleron rate is larger than the cosmological expansion rate at temperatures above the critical temperature and is smaller than the cosmological expansion rate at temperatures below the critical temperature. For each transition, B violation strongly takes place in the symmetric phase regions, but it essentially shuts off in the broken phase bubbles. This may provide B violation necessary for baryogenesis, as required by the first of Sakharov's conditions, in a relationship with nonequilibrium physics.

As summarized from our previous work \cite{phonglongvan}, both transitions in the EWPT sequence in this model are the first order, and they are sufficiently strong  (i.e., their strengths are larger than unity). Each transition proceeds through violent nucleation of bubbles of the broken phase as the Universe cools through the respective critical temperatures. This sequence of strong EWPTs may provide a source of large deviations from thermal equilibrium, as required by the third of Sakharov's conditions.

However, in order to establish that the RM331 model contains all necessary components for EWBG, we need to investigate the C- and CP- violating interactions in the model to ensure that the model satisfies the second of Sakharov's conditions. This is a focus in our next works.

\section*{Acknowledgment}
This research is funded by Vietnam  National Foundation for
Science and Technology Development (NAFOSTED)  under Grant No.
103.01-2014.51.
\\[0.3cm]

\end{document}